\documentclass[floats, prd, eqnum, showpacs, nofootinbib, 
twocolumn, 
eqsecnum]{revtex4-1}

\usepackage{color,graphicx}
\usepackage{amsfonts}
\usepackage{amssymb}
\begin{document}

\title{Comment on "Horizon-scale tests of gravity theories and fundamental physics from the Event Horizon Telescope image of Sagittarius A*"}
\author{Naoki Tsukamoto}\email{tsukamoto@rikkyo.ac.jp}

\affiliation{
Department of Physics, Faculty of Science, Tokyo University of Science, 1-3, Kagurazaka, Shinjuku-ku, Tokyo 162-8601, Japan \\
}

\begin{abstract}
Vagnozzi \textit{et al.} constrained the additional parameter of spacetimes with a photon sphere from the observation of the shadow of Sagittarius A* 
under the assumption that a distance to the Sagittarius A* and its mass parameter was estimated from other observations. 
They claimed that a Damour-Solodukhin wormhole with an additional parameter $\lambda$ is not an asymptotically-flat spacetime and that they gave the first 
robust observational constraint on the parameter $\lambda$ of the Damour-Solodukhin wormhole. However, they overlooked the fact that: 
(A) the Damour-Solodukhin wormhole spacetime is asymptotically flat, 
(B) the throat of the Damour-Solodukhin wormhole works as an effective photon sphere for $\lambda>\sqrt{2}/2$, and  
(C) not only a usual mass parameter but also the parameter $\lambda$ contributes the mass of the Damour-Solodukhin wormhole.
Because of the overlook (C), we realize that their constraint on the parameter $\lambda$ is invalid.
This is because their method corresponds to the following way:
They estimated the mass parameter from the other observations under the assumption $\lambda=0$ 
even though the value of the parameter $\lambda$ strongly affects the determination of the mass parameter, 
and then, they used the value of the mass parameter to constrain $\lambda$ from the observation of the shadow. 
We conclude that we should constrain the mass parameter and $\lambda$ from the shadow observation and other observations without the assumption $\lambda=0$.          
\end{abstract}

\maketitle

\section{Introduction}
In Ref.~\cite{Vagnozzi:2022moj}, Vagnozzi \textit{et al.} 
gave the constraints of an additional parameter of spacetimes with a photon sphere which is formed by unstable circular photon orbits 
from the observation of a black hole candidate which is called Sagittarius A* at the center of our galaxy observed by Event Horizon Telescope Collaboration~\cite{EventHorizonTelescope:2022wkp}
under the asusumption that the mass of the Sagittarius A* and its distance are given by other observations~\cite{GRAVITY:2020gka,EventHorizonTelescope:2019jan}.

Vagnozzi \textit{et al.} considered
the shadow of a Damour-Solodukhin wormhole~\cite{Damour:2007ap}. 
Its metric in coordinates ($\tilde{t}$, $r$, $\theta$, $\phi$) is given by
\begin{eqnarray}\label{eq:metric}
\mathrm{d}s^2
&=&-\left(1-\frac{2\tilde{M}}{r}+\lambda^2 \right) \mathrm{d}\tilde{t}^2+\frac{\mathrm{d}r^2}{1-\frac{2\tilde{M}}{r}} \nonumber\\
&&+r^2 \left( \mathrm{d}\theta^2+ \sin^2 \theta \mathrm{d}\phi^2 \right),
\end{eqnarray}
where $\tilde{M}$ is a mass parameter and $\lambda$ is the additional parameter.
It has a wormhole throat at $r=2\tilde{M}$. We assume the $Z_2$ symmetry of the wormhole spacetime against the throat.
We treat $\lambda$ as a free positive parameter on this paper in the same way as Ref.~\cite{Vagnozzi:2022moj}. 

At a spatial infinity $r \rightarrow \infty$, the metric~(\ref{eq:metric}) in the coordinates ($\tilde{t}$, $r$, $\theta$, $\phi$) becomes    
\begin{eqnarray}\label{eq:metric2}
\mathrm{d}s^2
&=&-\left(1+\lambda^2 \right) \mathrm{d}\tilde{t}^2+\mathrm{d}r^2
+r^2 \left( \mathrm{d}\theta^2+ \sin^2 \theta \mathrm{d}\phi^2 \right). \nonumber\\
\end{eqnarray}
The coordinates~($\tilde{t}$, $r$, $\theta$, $\phi$) are different from ones which are often used in a Minkowski spacetime.
Therefore, the physical interpretation of $\lambda$ and the constants of motion under the coordinates~($\tilde{t}$, $r$, $\theta$, $\phi$) can be uncertain for observers at far away from the wormhole or it would not be simple for them.

In Sec.~II, we comment on overlooks in Ref.~\cite{Vagnozzi:2022moj}:(A) its asymptotically flatness, (B) an effective photon sphere on its throat for $\lambda > \sqrt{2}/2$, and (C) the contribution of the parameter $\lambda$ to its mass.
We give a conclusion in Sec.~III. 
On this paper, we use the units in which a light speed and Newton's constant are unity.

\section{Comments on Vagnozzi \textit{et al.}~\cite{Vagnozzi:2022moj}}

\subsection{Asymptotically flatness}
Vagnozzi \textit{et al.}~\cite{Vagnozzi:2022moj} claimed that the wormhole is not asymptotically flat. 
This statement contradicts Ref.~\cite{Damour:2007ap} by Damour and Solodukhin.  
Here, we clearly show that it is asymptotically flat by transforming of the time coordinate $\tilde{t}$ 
into a new one $t$ defined by~\footnote{The new time coordinate used in Refs.~\cite{Bueno:2017hyj,Ovgun:2018fnk,Tsukamoto:2020uay}. Note that definition of the coordinate $t$ given by Eq.~(2.6) 
in Ref.~\cite{Tsukamoto:2020uay} has a typo 
and it should be read as Eq.~(\ref{eq:newt}) on this paper.}  
\begin{eqnarray}\label{eq:newt}
t=\sqrt{1+\lambda^2}\tilde{t}.
\end{eqnarray}
In the coordinates ($t$, $r$, $\theta$, $\phi$), the metric~(\ref{eq:metric}) is rewritten in  
\begin{eqnarray}\label{eq:metric3}
\mathrm{d}s^2
&=&-\left[1-\frac{2\tilde{M}}{(1+\lambda^2)r} \right] \mathrm{d}t^2+\frac{\mathrm{d}r^2}{1-\frac{2\tilde{M}}{r}} \nonumber\\
&&+r^2 \left( \mathrm{d}\theta^2+ \sin^2 \theta \mathrm{d}\phi^2 \right)
\end{eqnarray}
and it is asymptotically flat.~\footnote{
We keep to use the mass parameter~$\tilde{M}$. 
However, $\tilde{M}$ 
is often redefined as a new mass parameter $M$ by a relation~\cite{Bueno:2017hyj,Ovgun:2018fnk,Tsukamoto:2020uay} 
\begin{eqnarray}\label{eq:newM}
M = \frac{\tilde{M}}{1+\lambda^2}
\end{eqnarray}
and the metric~(\ref{eq:metric3}) is rewriten in 
\begin{eqnarray}\label{eq:metric5}
\mathrm{d}s^2
&=&-\left(1-\frac{2M}{r} \right) \mathrm{d}t^2+\frac{\mathrm{d}r^2}{1-\frac{2M\left(1+\lambda^2 \right)}{r}} \nonumber\\
&&+r^2 \left( \mathrm{d}\theta^2+ \sin^2 \theta \mathrm{d}\phi^2 \right).
\end{eqnarray}
} 
The mass parameter $\tilde{M}$ is an Arnowitt-Deser-Misner (ADM) mass in the line element~(\ref{eq:metric3}) as shown in Ref.~\cite{Tsukamoto:2020uay}.\footnote{The ADM mass $M(1+\lambda^2)$ was obtained as shown Eq.~(A6) in Ref.~\cite{Tsukamoto:2020uay}. It is equal to $\tilde{M}$ from Eq.~(\ref{eq:newM}) on this paper.}

\subsection{Effective photon sphere on the throat for $\lambda>\sqrt{2}/2$}
Vagnozzi \textit{et al.} obtained the shadow size, which is Eq. (38) in Ref.~\cite{Vagnozzi:2022moj}, as  
\begin{eqnarray}\label{eq:rsh}
r_{\mathrm{sh}}\simeq \frac{3\sqrt{3}}{1+\lambda^2} \tilde{M},
\end{eqnarray}
where we have recovered $\tilde{M}$, which is denoted by $M$ on Ref.~\cite{Vagnozzi:2022moj}, while they used units that the mass parameter is unity.
They used $\simeq$ because they misunderstood that the wormhole spacetime is not asymptotically flat. 
We should read Eq.~(\ref{eq:rsh}) as 
\begin{eqnarray}\label{eq:rsh2}
r_{\mathrm{sh}}= \frac{3\sqrt{3}}{1+\lambda^2} \tilde{M}.
\end{eqnarray} 
Equation~(\ref{eq:rsh2}) is equal to the critical impact parameter $b_\mathrm{m}$ of a ray under the coordinates~($t$, $r$, $\theta$, $\phi$) 
obtained by \"Ovg\"un~\cite{Ovgun:2018fnk} who is one of the authors of Ref.~\cite{Vagnozzi:2022moj}. See Eq.~(2.26) in Ref.~\cite{Ovgun:2018fnk}.~\footnote{
In Ref.~\cite{Ovgun:2018fnk}, $b_{m}$ is denoted by $u_m$. Note that $M$ on Ref.~\cite{Ovgun:2018fnk} and $\tilde{M}$ on this paper has the relation of Eq.~(\ref{eq:newM}) on this paper.
We notice that an effective photon sphere on the throat was overlooked in Ref.~\cite{Ovgun:2018fnk}.}  

However, Eq.~(\ref{eq:rsh2}) is correct only for $\lambda \leq \sqrt{2}/2$. 
We explain the reason why it is vaild only  for $\lambda \leq \sqrt{2}/2$ by following Ref.~\cite{Tsukamoto:2020uay}.
The Damour-Solodukhin wormhole spacetime has a photon sphere, which is formed unstable circular photon orbits, 
at $r=r_\mathrm{m}$. 
For $\lambda<\sqrt{2}/2$, $r_\mathrm{m}$ is defined by
\begin{eqnarray}\label{eq:photonsphere}
r_\mathrm{m}\equiv \frac{3\tilde{M}}{1+\lambda^2}.
\end{eqnarray}
For $\lambda=\sqrt{2}/2$, the radius $r_\mathrm{m}$
of the photon sphrere given by Eq.~(\ref{eq:photonsphere})
is equal to the radius of the throat, i.e., 
\begin{eqnarray}\label{eq:photonsphere2}
r_\mathrm{m}=2\tilde{M}.
\end{eqnarray}
In this case, the photon sphere on the throat is a kind of a marginally unstable photon sphere. 
For $\lambda>\sqrt{2}/2$, the throat works as an effective photon sphere. Thus, we define $r_\mathrm{m}$ as, for $\lambda>\sqrt{2}/2$,  
\begin{eqnarray}\label{eq:photonsphere3}
r_\mathrm{m}\equiv 2\tilde{M}.
\end{eqnarray}

A ray coming from spatical infinity falls into the throat for $b<b_\mathrm{m}$, 
where $b=b(r_0)$ is the positive impact parameter of the ray, 
where $r_0$ is the closest distance of the ray from the wormhole, 
and 
\begin{eqnarray}\label{eq:criticalimpactparameter}
b_\mathrm{m}\equiv b(r_0=r_\mathrm{m})
\end{eqnarray}
is the critical impact parameter of the ray.   
On the other hand, for $b>b_\mathrm{m}$, the ray coming from the spatial infinity is deflected by the wormhole 
and it goes back to the same spatial infinity. 
In the critical case with $b=b_\mathrm{m}$, the ray rotates around the (effective) photon sphere infinity times. 

The image angle of the (effective) photon sphere, which would be the observed angle of a ring image around the shadow, 
is given by $b_\mathrm{m}/D$, where $D$ is a distance between the wormhole and the observer.
If the distance is obtained from other observations, the size of the (effective) photon sphere is determined by the critical impact parameter $b_\mathrm{m}$.
For $\lambda \leq \sqrt{2}/2$, the critical impact parameter $b_\mathrm{m}$ is obtained as
\begin{eqnarray}\label{eq:criticalb1}
b_\mathrm{m}=\frac{3\sqrt{3} }{1+\lambda^2}\tilde{M}.
\end{eqnarray}
It is almost equal to the shadow size~(\ref{eq:rsh}) obtained by Vagnozzi \textit{et al.}~\cite{Vagnozzi:2022moj} and
it is equivalent with the critical impact parameter obtained by \"Ovg\"un~\cite{Ovgun:2018fnk}.
On the other hand, for $\lambda > \sqrt{2}/2$, it is given by~\cite{Tsukamoto:2020uay} 
\begin{eqnarray}\label{eq:criticalb2}
b_\mathrm{m}=\frac{2  \sqrt{1+\lambda^2}}{\lambda} \tilde{M}.
\end{eqnarray}
Thus, the shadow size~(\ref{eq:rsh}) obtained by Vagnozzi \textit{et al.}~\cite{Vagnozzi:2022moj} 
should be read as Eq.~(\ref{eq:criticalb2}) or 
\begin{eqnarray}
r_{\mathrm{sh}}=\frac{2  \sqrt{1+\lambda^2}}{\lambda} \tilde{M}
\end{eqnarray}
for $\lambda > \sqrt{2}/2$.

The throat of a rotating Damour-Solodukhin wormhole, which is often called Kerr-like wormhole, can affect its shadow in shape and size.  
In Ref.~\cite{Amir:2018pcu}, the effect of the throat of the rotating wormhole on the shadow was overlooked. 
Kasuya and Kobayashi modified the shadow of the rotating wormhole by taking an account for the effect of the throat~\cite{Kasuya:2021cpk}. 
See also Ref.~\cite{Shaikh:2018kfv} for the effect of a throat on a wormhole shadow.

\subsection{Contribution of $\lambda$ to its mass}
The coodinates ($t$, $r$, $\theta$, $\phi$) make the meaning of $\lambda$ clear.
From Eq.~(\ref{eq:metric3}), we realize that 
not only the mass parameter $\tilde{M}$ but also the parameter $\lambda$ contributes the mass of the Damour-Solodukhin wormhole.

However, Vagnozzi \textit{et al.}~\cite{Vagnozzi:2022moj} did not mention the effect of $\lambda$ to the mass 
and they assumed that the mass parameter $\tilde{M}$ could be estimated by other observations~\cite{GRAVITY:2020gka,EventHorizonTelescope:2019jan}.
The assumption is invalid since the value of $\lambda$ strongly affects the determination of $\tilde{M}$.
Thus, the constraint on $\lambda$ from the shadow is invalid.

\section{Conclusion}
In Ref.~\cite{Vagnozzi:2022moj}, Vagnozzi \textit{et al.} 
gave the constraint of an additional parameter of spacetimes with a photon sphere from the shadow observation under the assumption that 
the mass parameter $\tilde{M}$ is given by other observations. 
We should keep in mind that their constraint on the additional parameter is valid only if it does not contribute the mass of the spacetime.  
Unfortunately, they overlooked the contribution of the parameter $\lambda$ of the Damour-Solodukhin wormhole to its mass.  
We realize that they should not have estimated the mass parameter $\tilde{M}$ of the Damour-Solodukhin wormhole from the other
observations under the assumption $\lambda=0$ since $\lambda$ strongly affects the mass of the wormhole.
Thus, we conclude that their constraint on the parameter $\lambda$ is invalid. 
We suggest that they should give constraints on $\lambda$ and $\tilde{M}$ from the shadow observation and other observations without the assumption $\lambda=0$. 

On this paper, we have concentraced on the Damour-Solodukhin wormhole spacetime 
but we will give short comments on other spacetimes discussed in Ref.~\cite{Vagnozzi:2022moj}.
Their constraints on parameters in almost trivial spacetimes would be valid
but we should be careful to treat other non-trivial spacetimes discussed in Ref.~\cite{Vagnozzi:2022moj}. 
We should reconsider several non-trivial cases such as Eqs. (72), (92), (111), (129), and (132) 
because not only the usual mass parameter but also another parameter can contribute the mass of the spacetime.

\section*{Acknowledgements}
The author is grateful to Hideki Maeda for his comments on the asymptotically (locally) flatness and the ADM mass.

\end{document}